\documentclass[twocolumn]{aastex701}
\usepackage{graphicx}
\usepackage{amsmath}

\newcommand{\Alfven}{Alfv\'{e}n\,}
\newcommand{\Alfvenic}{Alfv\'{e}nic\,}

\begin{document}

\title{Hybrid Simulations of Proton Acceleration at Oblique High-$\beta$ Shocks}
\author[0009-0004-7654-0502]{Yevhen Kylivnyk}
\affiliation{The University of Chicago, 5640 S Ellis Ave, Chicago, IL 60637, USA}
\email{s}

\author[0000-0003-0939-8775]{Damiano Caprioli}
\affiliation{The University of Chicago, 5640 S Ellis Ave, Chicago, IL 60637, USA}
\affiliation{Enrico Fermi Institute, The University of Chicago, Chicago, IL 60637, USA}
\email{s}

\author[0000-0002-1879-457X]{Luca Orusa}
\affiliation{Department of Astronomy and Columbia Astrophysics Laboratory, Columbia University, 538 West 120th Street, New York,
NY 10027, USA}
\affiliation{Department of Astrophysical Sciences, Princeton University, Princeton, NJ 08544, USA}
\email{s}

\begin{abstract}
Collisionless shocks in the intracluster and intergalactic medium (ICM/IGM) are expected to energize both electrons and ions.
While electron acceleration is revealed by prominent radio emission, $\gamma$-ray emission from hadronic interactions remains undetected, suggesting that high-$\beta$ (ratio of thermal to magnetic pressure), low-Mach-number shocks cannot accelerate protons efficiently.
We present three-dimensional hybrid simulations of quasi-perpendicular
shocks with sonic Mach numbers $M_s\sim{3.4{-}17.5}$, focusing primarily on
high-$\beta$ conditions representative of cluster environments and including
several lower-$\beta$ comparison runs. We find that weak shocks with $M_s\lesssim5$ are consistently inefficient, with
$\varepsilon_{\rm CR}\lesssim0.1\%$, while inefficient cases can persist to
substantially higher $M_s$ depending on $M_A$ and $\beta$.
In contrast, stronger shocks ($M_s \gtrsim 10$) develop clear power-law tails with slopes $q \sim 4.0$ and reach $\varepsilon_{\rm CR} \sim 3\%$.
These results suggest that weak, oblique ICM shocks are generally unlikely to accelerate protons efficiently. However, reducing $\vartheta_{B_n}$ to $\sim 45^\circ$
leads to substantially higher acceleration efficiencies, indicating that magnetic obliquity plays a critical role in determining proton acceleration.
Our findings provide a microphysical framework for interpreting radio relic observations, whose polarization suggests that electrons are accelerated at oblique shocks, and the absence of cluster $\gamma$-ray detections.
\end{abstract}

\section{Introduction}
\label{sec:intro}

Collisionless shocks are ubiquitous in astrophysical plasmas,
arising in environments ranging from the solar wind and planetary bow shocks to supernova remnants (SNRs), galaxy clusters, active galactic nuclei (winds and jet lobes), and large-scale structure formation. In these systems, shocks convert bulk kinetic energy into thermal energy, turbulence, and nonthermal particles.

Strong shocks in SNRs are widely recognized as efficient cosmic ray (CR) accelerators via diffusive shock acceleration \citep[DSA, see][]{krymskii77, axford+78, bell78a, bell78b, blandford+78}. These shocks typically occur in plasmas characterized by high sonic and \Alfvenic Mach numbers, defined as the ratio between the shock velocity and local sound and \Alfven speed respectively, conditions that favor both magnetic-field amplification and efficient injection of thermal particles.
In contrast, the situation is far less clear for \emph{weak shocks}, such as those pervading the intracluster and intergalactic medium (ICM/IGM).

In galaxy clusters, which are the largest gravitationally bound structures in the Universe, shocks naturally arise during hierarchical assembly and merger events. As clusters collide, the ICM is driven to supersonic velocities, generating large-scale shock waves that propagate through the cluster medium \citep[see][for a review]{brunetti+14}.
Such shocks are characterized by modest sonic Mach numbers ($M_s \lesssim 5$) and plasma $\beta\gg 1$, with $\beta \equiv P_{\rm th}/P_B$ defined as the ratio of thermal to magnetic pressures, a regime where the microphysics of particle injection and acceleration remain poorly understood.

Observations of cluster-scale radio relics demonstrate that relativistic electrons can indeed be energized at such shocks \citep[e.g.,][]{ferrari06, vanweeren+10,brunetti+14,vanweeren+19}, and the direction of the inferred polarization suggests that the radio-bright regions are predominantly quasi-perpendicular \citep[e.g.,][]{bonafede+09, vanweeren+12, Jones+21}.
Modern kinetic particle-in-cell (PIC) plasma simulations present \textit{hints} of electron acceleration in high-$\beta$ oblique shocks \citep[e.g.,][]{guo+14a, guo+14b, ha+18, ha+20, ha+23, kobzar+21}, though clear-cut DSA tails have never been reported in the literature.

\begin{figure*}[ht!!]
\centering
\includegraphics[width=1.\textwidth,clip=true,trim= 0 100 00 80]{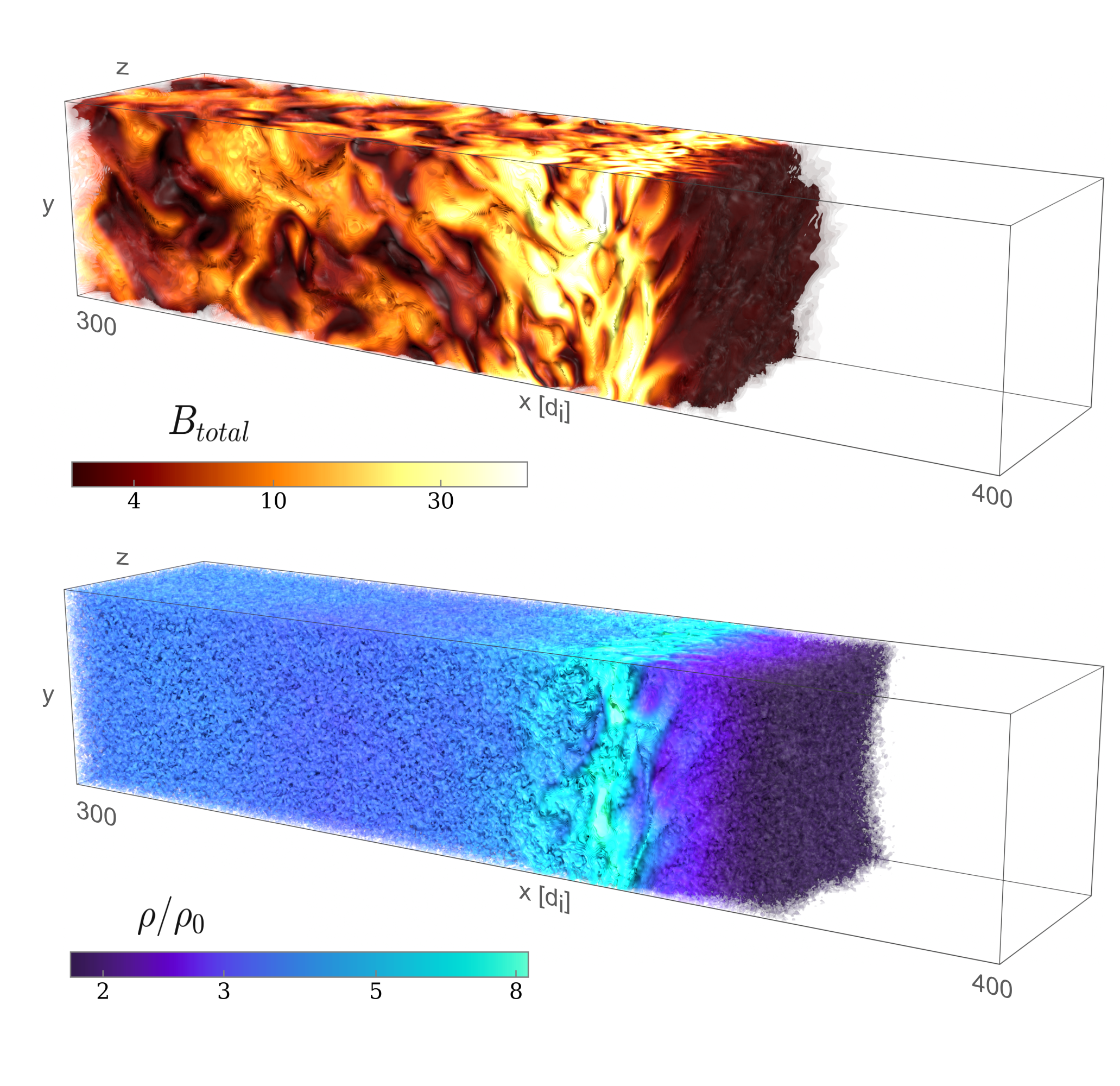}
\caption{3D rendering of total magnetic field $B_{total}$ and density (top and bottom panel, respectively) for Run 12 in Table \ref{tab:params}.}
\label{fig: 3Ds snapcshot}
\end{figure*}

In any case, the electron acceleration efficiency inferred from simulations \citep[e.g.,][]{park+15, shalaby+22, gupta+25, gupta+24b} or extrapolated from typical SNR shocks \citep[e.g.,][]{berezhko+04a, morlino+12, slane+14, mcginness+26} are typically too low to account for the observed emission \citep{ferrari06, vanweeren+10, brunetti+14, vanweeren+19}.
Moreover, the observed radio spectral indices are difficult to reconcile with the relatively low Mach numbers inferred from X-ray observations \citep[e.g.,][]{ogrean+13, vanweeren+19, whittingham+26}.
This discrepancy raises the possibility that pre-existing seed populations, re-acceleration processes \citep[e.g.,][]{pinzke+10,pinzke+13, kang+10}, or shock obliquity effects play a critical role.
Meanwhile, whether \textit{protons} are accelerated efficiently at high-$\beta$ cluster shocks remains a matter of active debate, with important implications for the expected level of $\gamma$-ray emission from hadronic interactions, which has so far eluded detection by \emph{Fermi}-LAT \citep[e.g.,][]{Ackermann+14, Ackermann+16, vazza+16, wittor+20}.
This contrast between efficient electron acceleration and suppressed proton injection seems to be a key feature of high-$\beta$ shocks.

Numerical simulations provide a powerful tool to probe these regimes. Full-PIC studies can capture both electron and ion microphysics, but are computationally prohibitive for the spatial and temporal scales required to follow proton acceleration at cluster shocks. Hybrid approaches \citep{winske+03, lipatov02}, which treat ions kinetically while modeling electrons as a charge-neutralizing fluid, offer a more tractable way to investigate ion injection, wave excitation, and long-term ion acceleration processes.

Early numerical studies of weak, high-$\beta$, quasi-perpendicular collisionless shocks relevant to the ICM, based on both full-PIC and hybrid simulations, generally found inefficient ion injection and little evidence for substantial nonthermal ion populations \citep[e.g.,][]{amano+09a, guo+14a, guo+14b, caprioli+14a, ha+18, xu+20, kumar+21, bohdan+21}. These investigations were initially performed in one-dimensional (1D) and 2D setups, while only a limited number of studies have explored fully three-dimensional configurations \citep[e.g.,][]{matsumoto+17, boula+24}. However, due to the high computational cost of kinetic simulations, existing 3D studies have generally been restricted to relatively small transverse domains and short evolutionary timescales, limiting their ability to capture long-term particle acceleration and large-scale shock corrugation effects.

\citet{boula+24} focused on high-$\beta$ shocks, similar to those considered here, studying the magnetic structures at the shock front in 2D hybrid simulations spanning different $M_A$, $M_s$ and angles, and also via two 3D simulations with $M_s=3$ and $M_A=6$ and $8.6$. Particle acceleration was not investigated in
that work. More recently, self-consistent 3D hybrid simulations have demonstrated that efficient ion acceleration can arise naturally at low-$\beta$ and high $M_A$, quasi-perpendicular shocks characteristic of the interstellar medium (ISM)
\citep{orusa+23,orusa+25a,orusa+25b}.
The 3D dynamics is crucial to capture the cross-field diffusion of ions in oblique shocks \citep[][]{jones+98}, and this motivates this paper's choice of focusing on 3D runs.
In this work, we use hybrid simulations to extend this analysis to quasi-perpendicular (magnetic obliquity $\vartheta_{B_n}\gtrsim 60^\circ$) shocks with $M_s \sim 3.4{-}17.5$, focusing primarily on high-$\beta$ conditions and including several lower-$\beta$ comparison runs, representative of conditions in the ICM and IGM. We focus on quantifying proton acceleration efficiency and exploring how it depends on shock parameters such as Mach number, obliquity, and plasma $\beta$. By systematically investigating this regime, we aim to clarify whether cluster shocks can contribute significantly to the nonthermal energy budget and to provide a microphysical basis for interpreting radio relic observations.

The paper is organized as follows. Section~\ref{sec:method} describes the numerical setup. Section~\ref{sec:results} presents the dependence of proton acceleration on the shock parameters and introduces an empirical fit for the acceleration efficiency. Section~\ref{sec:Discussion} discusses the astrophysical implications of our findings, and Section~\ref{sec:conclusions} summarizes our conclusions.

\section{Numerical Setup}
\label{sec:method}

Our simulations employ the 3D hybrid code \texttt{dHybridR} \citep{haggerty+19a}, in which ions are treated kinetically while electrons are a charge-neutralizing fluid. Shocks are generated by injecting a supersonic upstream flow with velocity $v_{0}$ (measured in the downstream rest frame) along the $-x$ direction against a reflecting wall, producing a shock that propagates in the $+x$ direction into a background magnetic field $B_0$ that makes an angle $\vartheta_{B_n}$ with the $x$-axis.

All lengths are normalized to the upstream ion skin depth,
$
d_i \equiv \frac{c}{\omega_{p}},
$
where $c$ is the speed of light and
$
\omega_{p} \equiv \sqrt{4\pi n_0 e^2/m}
$
is the upstream ion plasma frequency, with $n_0$ the upstream ion number density, $e$ the elementary charge, and $m$ the proton mass.
Time is normalized to the inverse ion cyclotron frequency,
$
\Omega_{c}^{-1} \equiv\frac{m c}{e B_0},
$
where $B_0$ is the upstream magnetic field strength. Velocities are expressed in units of the upstream \Alfven speed,
$
v_A \equiv B_0/\sqrt{4\pi n_0 m},
$
and energies are scaled to
\[
E_{\rm sh} \equiv \frac{1}{2} m v_{0}^2.
\]
The sonic Mach number is defined as
$M_s \equiv v_{sh}/c_s$, where $v_{sh}$ is the shock front speed in the upstream rest frame, $c_s$ is the upstream sound speed,
and the Alfv\'enic Mach number as
$M_A \equiv v_{sh}/v_A$.
The corresponding Mach number in the downstream (simulation) rest frame is slightly smaller and given by a Galilean transformation that depends on the shock compression ratio
$r$ as $M_{s,\rm sims}=M_s\,(r-1)/r$.

\begin{figure*}[ht!!]
\centering
\includegraphics[width=1.\textwidth]{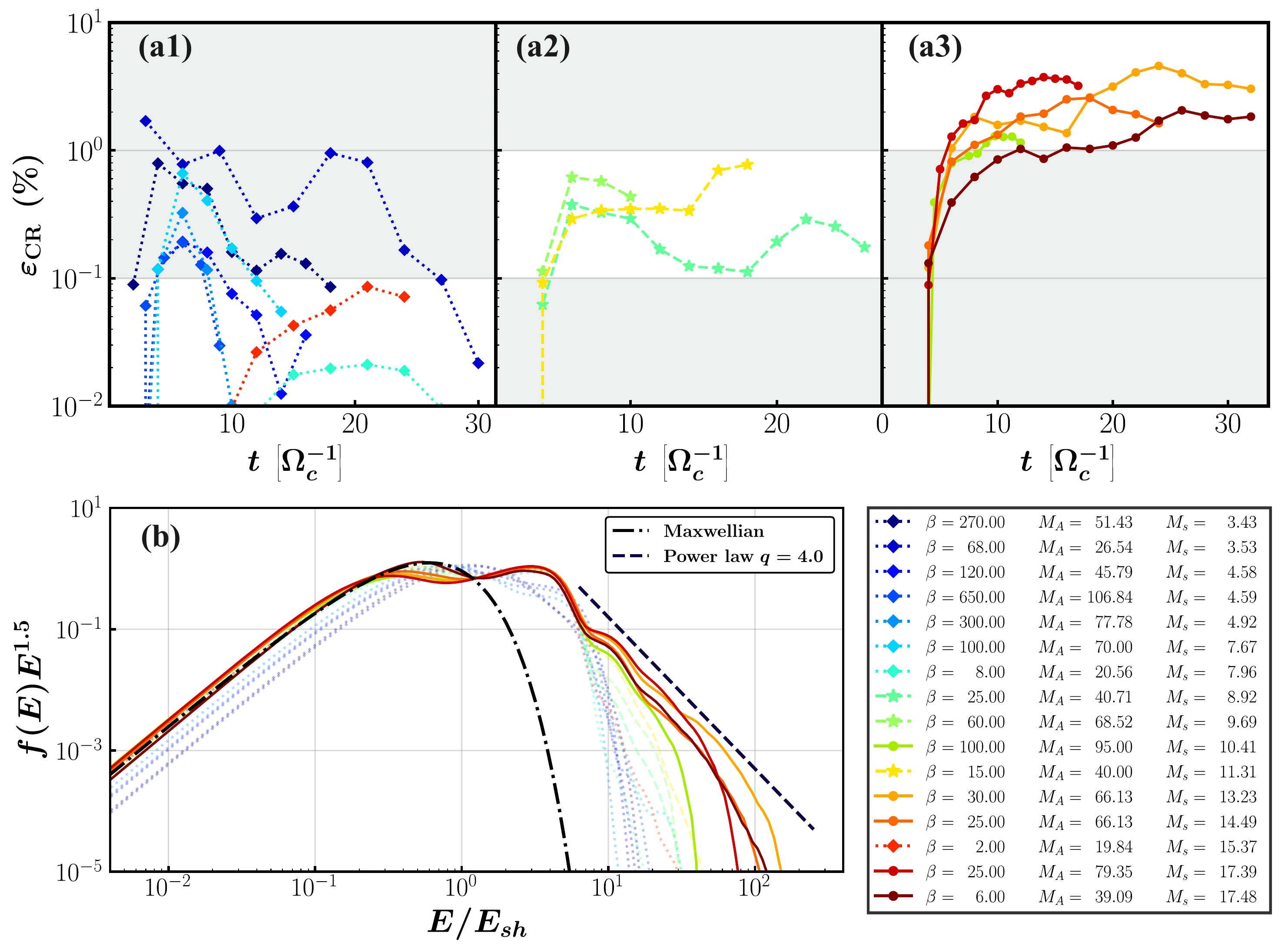}
\caption{\textbf{(a)} CR acceleration efficiency, $\varepsilon_{\rm CR}$, as a function of time. \textbf{(a1)} corresponds to highly inefficient shocks with $\varepsilon_{\rm CR}<0.1\%$, \textbf{(a2)} to intermediate efficiencies with $0.1\%\leq\varepsilon_{\rm CR}<1\%$, and \textbf{(a3)} to efficient shocks with $\varepsilon_{\rm CR}\geq1\%$.
\textbf{(b)} Energy spectra at the final timestep for selected runs. For highly efficient cases, Maxwellian and power-law fits of the form $f(E)\propto E^{-q}$ are shown (black lines), where $q$ is the spectral slope. Shocks with higher $M_s$ develop a prominent nonthermal tail, while low-$M_s$ shocks remain nearly thermal. All the simulations reported in this plot are performed at $\vartheta_{B_n}=80^\circ$.}
\label{fig:Last_time_spectra}
\end{figure*}

The simulation domain spans $L_x \times L_y \times L_z$ in units of the ion skin depth, $d_i$. The transverse dimensions are fixed at $L_y=L_z=20\,d_i$, while the longitudinal size varies between $L_x\sim300$ and $800\,d_i$, depending on the run. In all simulations, the box is sufficiently large to prevent accelerated particles from reaching the upstream boundary over the simulated timescale. We adopt a spatial resolution of 8 cells per $d_i$ and 8 macroparticles per cell; the validity of this choice is discussed in Appendix~\ref{appendix:resolution}.
This resolution is higher than that typically adopted in the literature for $\beta\sim1$ shocks \citep{caprioli+14a} and was chosen to avoid numerical artifacts, which are more prominent in high-$\beta$ plasmas.

Table~\ref{tab:params} summarizes all runs, covering a range of $\beta = \frac{2}{\gamma}\left(\frac{M_A}{M_s}\right)^2\sim2{-}650$ (where $\gamma = 5/3$ is the gas adiabatic index) and $M_s\sim3.4{-}17.5$. Unless otherwise stated, $\vartheta_{B_n}=80^\circ$.

\begin{table}[t]
\caption{Summary of 3D hybrid simulations of
$\vartheta_{B_n}=80^\circ$ shocks.
For each run, we report the input plasma parameters, the measured shock
compression ratio $r$ at the final timestep, and the cosmic-ray
acceleration efficiency $\varepsilon_{\rm CR}$.
Following Figure~\ref{fig:efficiency}, the efficiencies are classified
as low ($\varepsilon_{\rm CR}<0.1\%$), intermediate (
$0.1\%\leq\varepsilon_{\rm CR}<1\%$), or high (
$\varepsilon_{\rm CR}\geq1\%$). For runs in the
high-efficiency regime, we report the time-averaged efficiency over the last
eight timesteps and its standard deviation.}
\label{tab:params}

\centering
\renewcommand{\arraystretch}{1.3}
\begin{tabular}{ccccc|cc}
\hline
\hline
Run $\#$ & $\beta$ & $M_s$ & $M_A$ & $t_{\rm end}\,[\Omega_c^{-1}]$ & $r$ & $\varepsilon_{\rm CR}$ (\%) \\
\hline
1  & 270 & 3.43  & 51.4  & 18.0 & 2.4 $\pm$ 0.3 & $<0.1$ \\
2  & 68  & 3.53  & 26.5  & 30.0 & 2.3 $\pm$ 0.3 & $<0.1$ \\
3 & 120 & 4.58 & 45.8 & 18.0 & $2.9 \pm 0.4$ & $<0.1$ \\
4  & 650 & 4.59  & 106.8 & 9.0  & 2.9 $\pm$ 0.4 & $<0.1$ \\
5  & 300 & 4.92  & 77.8  & 8.0  & 2.8 $\pm$ 0.3 & $<0.1$ \\
6  & 100 & 7.67  & 70.0  & 10.0 & 3.5 $\pm$ 0.2 & $<0.1$ \\
7  & 8   & 7.96  & 20.6  & 27.0 & 3.7 $\pm$ 0.1 & $<0.1$ \\
8  & 25  & 8.92  & 40.7  & 26.0 & 3.8 $\pm$ 0.1 & 0.1--1 \\
9  & 60  & 9.69  & 68.5  & 10.0 & 3.7 $\pm$ 0.1 & 0.1--1 \\
10 & 100 & 10.41 & 95.0  & 12.0 & 3.8 $\pm$ 0.2 & $1.1 \pm 0.2$ \\
11 & 15  & 11.31 & 40.0  & 18.0 & 4.0 $\pm$ 0.1 & 0.1--1 \\
12 & 30  & 13.23 & 66.1  & 32.0 & 4.1 $\pm$ 0.1 & $3.5 \pm 0.6$ \\
13 & 25  & 14.49 & 66.1  & 24.0 & 4.1 $\pm$ 0.1 & $2.0 \pm 0.4$ \\
14 & 2   & 15.37 & 19.8  & 24.0 & 4.1 $\pm$ 0.1 & $<0.1$ \\
15 & 25  & 17.39 & 79.4  & 10.0 & 4.1 $\pm$ 0.1 & $3.3 \pm 0.3$ \\
16 & 6   & 17.48 & 39.1  & 10.0 & 4.3 $\pm$ 0.1 & $1.6 \pm 0.4$ \\
\hline
\hline
\end{tabular}
\end{table}
The simulation duration was chosen individually for each run based on the
computational cost and the time required to determine whether a persistent
nonthermal ion population developed. Runs that rapidly settled into a
low-efficiency state, or whose acceleration regime could already be clearly
identified, were terminated earlier, whereas runs exhibiting continued
evolution of the nonthermal component were followed for longer times.
Because $\varepsilon_{\rm CR}$ can fluctuate, particularly in inefficient
runs such as run~1, we do not interpret small variations in its instantaneous
final value as physically significant. Instead, such cases are classified
according to their acceleration-efficiency regime, with
$\varepsilon_{\rm CR}<0.1\%$ regarded as inefficient.

\section{Results}
\label{sec:results}

\subsection{Shock structure}
\label{subsec: shock structure}

Figure~\ref{fig: 3Ds snapcshot} shows the three-dimensional structure of the total magnetic-field strength and ion density for Run~12 at the final simulation time.
The upstream plasma is located to the right of the shock, while the compressed downstream region extends to the left.
The shock transition is marked by a sharp increase in both density and magnetic-field strength.
Rather than remaining planar, the shock front develops pronounced corrugations across the transverse directions, producing spatial variations in the local compression and magnetic-field amplification.

Behind the shock, the magnetic field develops an inhomogeneous, filamentary structure, with characteristic transverse scales of approximately $10\,d_i$ (see \citep[][]{orusa+26}).
Regions of enhanced magnetic-field strength broadly coincide with regions of increased density, indicating that the downstream magnetic structure is associated with compression and deformation of the plasma at the corrugated shock front.
The strongest magnetic amplification occurs close to the shock transition, while farther downstream the field remains turbulent and spatially intermittent.

The degree of shock structuring depends strongly on the shock strength.
At high $M_s$, the shock develops substantial corrugation and downstream magnetic turbulence, creating spatially intermittent regions of low and high magnetic-field strength.
Such three-dimensional structure facilitates cross-field transport and allows a fraction of the suprathermal ions to return toward the shock, supporting continued acceleration.
In contrast, low-$M_s$ shocks relax toward a comparatively laminar state dominated by simple downstream magnetic compression.
The absence of strong fluctuations and low-$B$ channels prevents particles advected downstream from returning to the shock efficiently, thereby suppressing injection and subsequent acceleration \citep[][]{orusa+25b}.

\subsection{Acceleration efficiency}
\label{subsec:Acceleration efficiency}

We quantify the acceleration efficiency, $\varepsilon_{\rm CR}$, as the fraction of the downstream ion energy contained in nonthermal particles. Following standard practice \citep[e.g.,][]{caprioli+14a, johlander+21}, we define
\begin{equation}
\varepsilon_{\rm CR} \equiv
       \int_{E_{\rm inj}}^{\infty} E f(E)\, dE \Big/ \int_{0}^{\infty} E f(E)\, dE,
\label{eq:efficiency}
\end{equation}
where $f(E)$ is the downstream ion energy distribution. We adopt $E_{\rm inj} \simeq 10 E_{\rm sh}$ as the threshold separating the thermal and suprathermal populations from the proper DSA one \citep{caprioli+15}.

The value of $\varepsilon_{\rm CR}$ is calculated within a downstream region of width $\sim100\,d_i$ behind the shock front, though varying the integration interval may lead to $10{-}20\%$ relative corrections. For runs in the high-efficiency regime, we report the efficiency averaged over the last eight timesteps, with the standard deviation over this interval quoted as the uncertainty.

Figure~\ref{fig:Last_time_spectra}a(1--3) shows the time evolution of $\varepsilon_{\rm CR}$ for runs grouped by their late-time acceleration efficiencies: $\varepsilon_{\rm CR}<0.1\%$ (a1), $0.1\%\leq\varepsilon_{\rm CR}<1\%$ (a2), and $\varepsilon_{\rm CR}\geq1\%$ (a3). We call these regimes low, intermediate, and high acceleration efficiencies, respectively.
Figure~\ref{fig:Last_time_spectra}b shows the downstream particle spectra at the final timestep for a representative subset of runs. For $M_s \gtrsim 10$, a power-law tail emerges with energy slope $q \sim 4.0$, much steeper than the standard DSA prediction of $q \sim 1.5$ for non-relativistic particles, but consistent with previous results for perpendicular and quasi-perpendicular shocks at moderate Mach numbers \citep{orusa+23, orusa+25b}. In contrast, shocks with $M_s \sim 3{-}9$ exhibit only a weak suprathermal excess. These suprathermal ions are produced primarily by reflection and shock-drift acceleration (SDA) at the shock transition, where they gain energy while drifting along the shock front in the motional electric field. However, most are subsequently advected downstream rather than injected into sustained DSA. For $M_s\lesssim5$, even this suprathermal component remains negligible.

\begin{figure}[t!]
\centering
\includegraphics[width=1.\columnwidth]{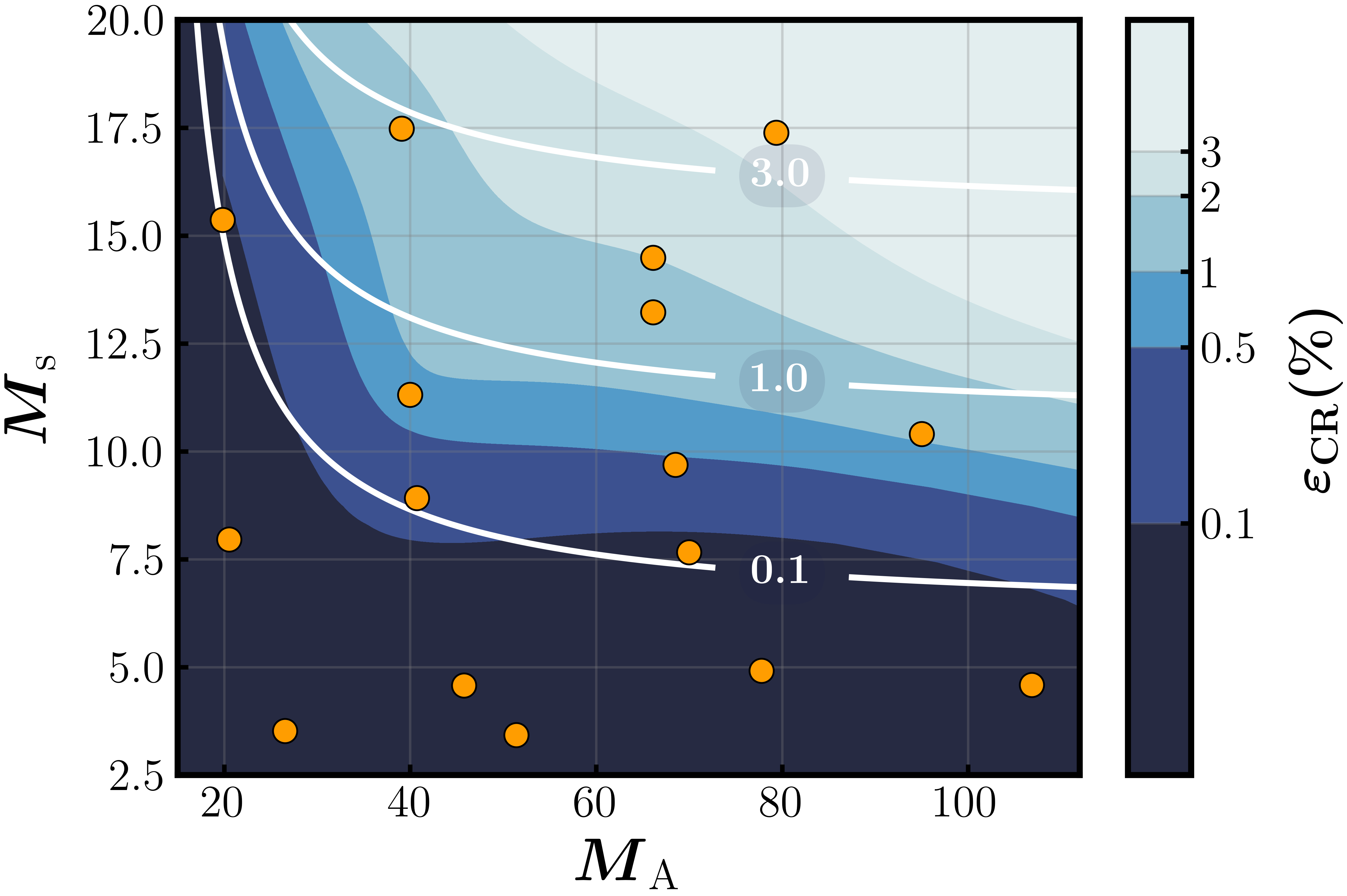}
\caption{Asymptotic CR acceleration efficiency, $\varepsilon_{\rm CR}$, as a function of the Alfv\'enic and sonic Mach numbers ($M_A$ and $M_s$), for $\vartheta_{B_n}=80^\circ$. Points denote individual simulation runs. Perpendicular shocks with low $M_s$ remain inefficient throughout the evolution. White lines indicate empirical fits for different values of $\varepsilon_{\rm CR}$ (Equation \ref{eq:fit}).}
\label{fig:efficiency}
\end{figure}

Figure~\ref{fig:efficiency} shows $\varepsilon_{\rm CR}$ for a range of plasma $\beta$ values and sonic Mach numbers. We find that low-$M_s$ perpendicular shocks ($M_s \lesssim 5$) are rather inefficient, with $\varepsilon_{\rm CR} \lesssim 0.1\%$ at late times.
In contrast, higher-$M_s$ shocks ($M_s \gtrsim 10$) can reach efficiencies as high as $\varepsilon_{\rm CR}\sim3.5\%$, indicating the development of an appreciable nonthermal ion component.
To facilitate inclusion of these results into large-scale simulations of galaxy clusters \citep[e.g.,][]{vazza+16, kang+13, ryu+19, dominguez+19, boess+24}, we provide an empirical fit to the measured efficiencies as a function of $M_s$ and $M_A$ (contours in Figure \ref{fig:efficiency}):

\begin{equation}\label{eq:fit}
\varepsilon_{\rm CR}(M_s,M_A)
= 0.024
\left(
M_s - 4.1 - \frac{70}{M_A - 12}
\right)^2.
\end{equation}
This fit is valid for $\vartheta_{B_n}=80^\circ$.
We set $\varepsilon_{\rm CR}=0$ wherever the efficiency predicted by Equation~\ref{eq:fit} falls below $0.1\%$.
We also stress that Equation~\ref{eq:fit} is an empirical interpolation calibrated only over the range explored by our simulations,
$
\left[ 3.5 \lesssim M_s \lesssim 17.5, \ \ 19 \lesssim M_A \lesssim 107\right].
$

In the low-$M_s$ regime, particle acceleration remains strongly suppressed regardless of $M_A$ or plasma $\beta$, indicating that $M_s$ is the primary parameter controlling both the acceleration efficiency and the spectral shape.

Low-$M_s$ shocks relax into a laminar state characterized by simple magnetic compression downstream and an absence of the strong turbulence required for the porosity-driven acceleration mechanism identified by \citet{orusa+23,orusa+25b}. As a result, particle injection remains inefficient: particles are efficiently advected downstream and cannot return upstream due to the absence of scattering centers or low-$B$ channels, leading to suppressed particle acceleration.

Since throughout the range of parameters considered ion acceleration remains modest ($\lesssim3.5\%$), we do not find strong modifications of the shock dynamics \citep{haggerty+20, haggerty+26}, with the measured compression ratios close to those expected from Rankine–Hugoniot relations.

\subsection{Angle dependence}
\label{subsec:Angle dependence}

\begin{figure}[ht]
\centering
\includegraphics[width=1.\columnwidth]{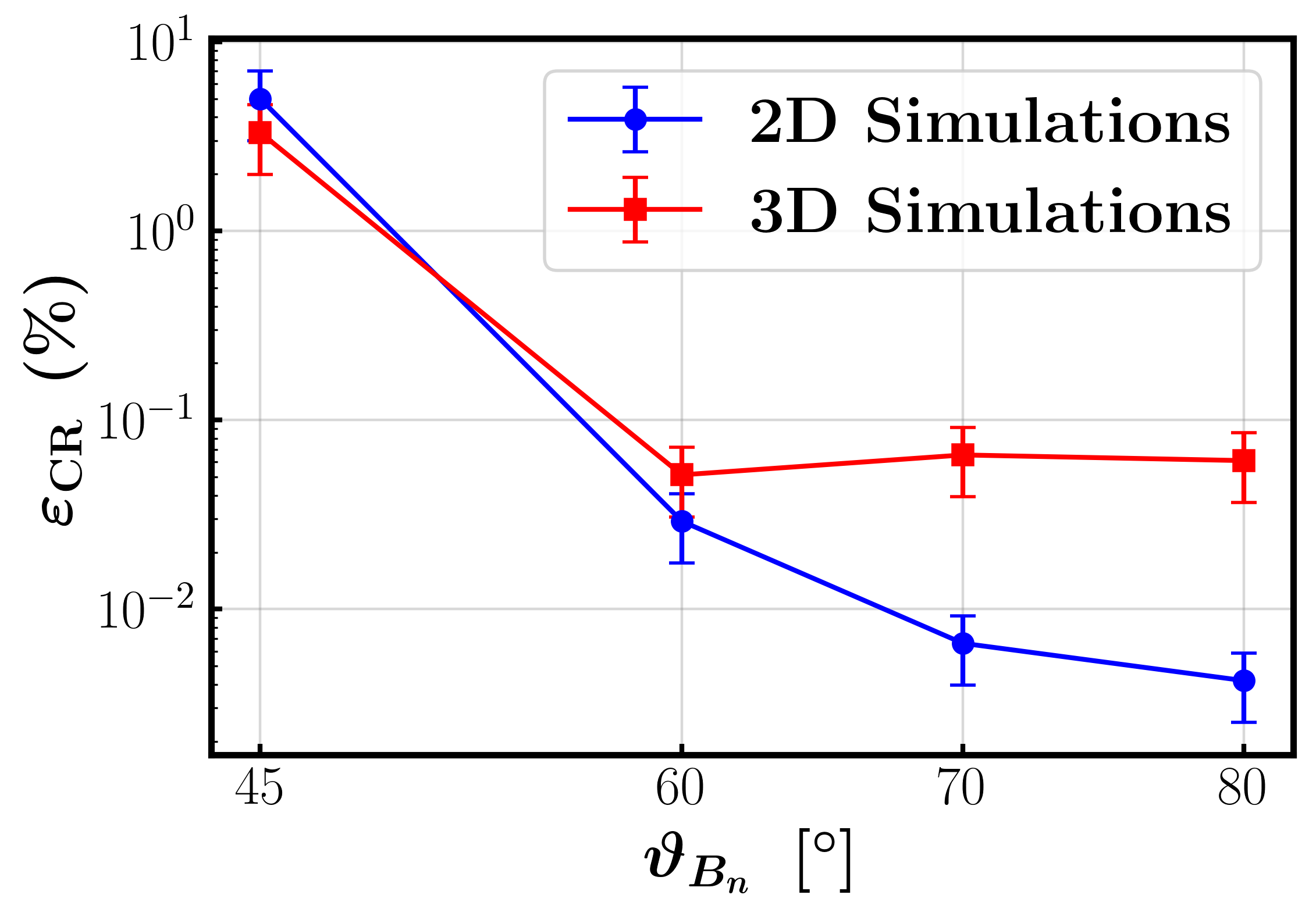}
\caption{Acceleration efficiency measured at the final simulation time, $\varepsilon_{\rm CR}$, as a function of shock obliquity, $\vartheta_{B_n}$. The final times are $t_{\rm last} = [160.0,\,128.0,\,128.0,\,128.0]\,\Omega_{c}^{-1}$ for the 2D runs and $[64.0,\,36.0,\,18.0,\,18.0]\,\Omega_{c}^{-1}$ for the 3D runs, ordered by increasing $\vartheta_{B_n}$.
All runs adopt the plasma parameters of Run~3 in Table~\ref{tab:params}. At large obliquities, both configurations are inefficient (especially in 2D) and converge to typical quasi-parallel values of $\varepsilon_{\rm CR}\simeq 10\%$ for $\vartheta_{B_n}\lesssim 45^\circ$.}
\label{fig:angle_eff}
\end{figure}

Since quasi-perpendicular shocks in our parameter range exhibit very low acceleration efficiency, we now investigate at which obliquity particle injection becomes more effective, and for which inclinations 3D simulations are required.

To this end, we perform a dedicated set of 2D and 3D simulations at fixed high $\beta$ and low $M_s$ from run $\#$ 3 (see Table \ref{tab:params}), i.e., conditions under which perpendicular shocks are inefficient.
The simulations are evolved up to final times of $t_{\rm last} = [160.0,128.0,128.0,128.0]\,\Omega_{c}^{-1}$ for the 2D runs and $[64.0,36.0,18.0,18.0]\,\Omega_{c}^{-1}$ for the 3D runs, ordered by increasing obliquity $\vartheta_{B_n} = [45^\circ, 60^\circ, 70^\circ, 80^\circ]$.
For highly oblique configurations, SDA sets in promptly after shock formation, so the presence (or absence) of accelerated particles can be assessed at early times, without the need for long integrations.
In contrast, quasi-parallel configurations require significantly longer runtimes to reach comparable evolutionary stages. While the $45^\circ$ cases in both 2D and 3D are not fully saturated, they still allow us to draw robust conclusions on whether decreasing the obliquity enhances the efficiency of particle acceleration.

Figure~\ref{fig:angle_eff} summarizes the dependence of $\varepsilon_{\rm CR}$ on the obliquity angle $\vartheta_{B_n}$ at the final timestep reached for the different simulations.
As expected \citep[][]{caprioli+14a}, acceleration efficiency increases as the shock becomes more quasi-parallel (i.e., lower $\vartheta_{B_n}$). At high obliquities, both 2D and 3D shocks are inefficient ($\varepsilon_{\rm CR} < 0.1\%$), although the 3D runs systematically achieve higher efficiencies than their 2D counterparts.
As the obliquity decreases, the efficiency rises, and both geometries converge to comparable values around $\vartheta_{B_n} \sim 45^\circ$. This trend is consistent with previous numerical studies \citep[e.g.,][]{orusa+23,caprioli+14b}, and suggests that for $\vartheta_{B_n} \lesssim 50^\circ$ 2D simulations are well suited to study ion acceleration in a computationally efficient way.
\section{Discussion}
\label{sec:Discussion}

In the context of the ICM, our results carry important phenomenological implications.
All simulations presented here probe shocks with $M_s \gtrsim 3.4$, where the quoted Mach numbers are measured in the upstream frame. These values significantly exceed the values typically inferred for cluster merger shocks, $M_s \simeq 1{-}1.5$ \citep[e.g.,][]{vanweeren+19}.

Running lower-$M_s$ shocks would require a setup different from the reflecting wall, but --- since our simulations show that even $M_s \sim 3.5{-}6$ quasi-perpendicular shocks fail to inject ions efficiently --- a natural extrapolation is that weaker shocks should be even less capable of producing nonthermal ions.
This expectation is further supported by the monotonic scaling of injection efficiency with Mach number found in hybrid-PIC and semi-analytic studies \citep[e.g.,][]{caprioli+14a, kang+13b}.
Our results therefore suggest that quasi-perpendicular ICM shocks with $M_s \lesssim 2$ contribute negligibly to the CR ion population.

This conclusion does not necessarily extend to quasi-parallel shocks, for which efficient ion acceleration has been reported even at relatively low Mach numbers \citep[e.g.,][]{caprioli+14a}, with recent dedicated studies of high-$\beta$ ICM-like shocks also indicating substantial ion acceleration efficiencies (\cite{ly+26}, Sharma \& Caprioli, in preparation).
However, the polarization of radio relics points to highly oblique regions to be synchrotron bright, as seen for instance in the Sausage and Toothbrush relics \citep[e.g.,][]{vanweeren+10, vanweeren+12}, which strongly suggest that electron acceleration does not follow the same trend as ion DSA.

Indeed, tensions between radio and $\gamma$-ray observations are commonly discussed assuming an electron-to-proton ratio comparable to supernova remnants, $K_{ep}\sim10^{-3}$. If ion acceleration is strongly suppressed at quasi-perpendicular ICM shocks, as suggested by our simulations, then the effective $K_{ep}$ could be substantially larger, formally diverging in the limit of vanishing ion acceleration.

An additional implication concerns the spectral slopes of the accelerated ions.
Even in the simulations that develop clear nonthermal tails, we find energy spectra with $q\sim4.0$, significantly steeper than the canonical DSA prediction.
Such steep spectra imply that a smaller fraction of the nonthermal energy is carried by the highest-energy ions, further suppressing the expected hadronic $\gamma$-ray emission.
Therefore, not only is ion acceleration inefficient at quasi-perpendicular ICM shocks, but the accelerated particles also exhibit softer spectra than expected from standard DSA.

It would be desirable to run hybrid simulations of very low-$M_s$ shocks, though properly capturing energy dissipation for $M_s\lesssim 2$ may involve resistivity and electron physics, especially at subcritical shocks \citep[e.g.,][]{balogh+13}.
Moreover, the canonical setup with a piston or a reflecting wall naturally produces an initial superposition of two flows, hence an effective density contrast of $\sim 2$, which would match the compression ratio of a shock with at least $M_s\sim 2$.
Running simulations in the transonic regime $M_s\lesssim 1.5$ thus requires a different driving, and should be done in full PIC to also address electron acceleration, which we defer to future works.

Finally, it is important to comment on the possible role of pre-existing turbulence, not included in the simulations presented here.
The ICM is prone to a variety of micro-instabilities, such as mirror and firehose instabilities \citep[e.g.,][]{kunz+14a, schekochihin+09}, which may generate fluctuations important for ion scattering \citep[][]{reichherzer+25, diesing+25b}.
Our simulations, initialized with a laminar upstream plasma, do not include these effects and therefore probe the ``clean'' injection properties of shocks.
Introducing controlled upstream perturbations in future studies, similarly to what was done by \citet{karol+23} for low-$\beta$ plasmas, may help determine whether weak shocks could achieve non-zero acceleration efficiency when embedded in a turbulent ICM environment.
Finally, reacceleration of pre-existing CR (both electrons and ions) is expected to be effective also for quasi-perpendicular configurations \citep[][]{caprioli+18}, and should not be ruled out as a potential way of tapping into the large energy reservoir of oblique ICM shocks.

\section{Summary and Conclusions}
\label{sec:conclusions}

We used 3D hybrid simulations to investigate proton acceleration at oblique collisionless shocks under conditions representative of the ICM and IGM, characterized by high plasma $\beta$ and low sonic Mach numbers.

Our results reveal an apparent Mach-number threshold for efficient proton acceleration for $\vartheta_{B_n}=80^\circ$.
Shocks with $M_s \lesssim 4.9$ produce negligible suprathermal populations ($\varepsilon_{\rm CR} \lesssim 0.1\%$; see Figure~\ref{fig:Last_time_spectra}).
In contrast, several stronger shocks with $M_s \gtrsim 10$ develop extended power-law tails that grow with time, but with spectral slopes $q \sim 4.0$, significantly steeper than the canonical DSA prediction;
in this regime the acceleration efficiencies can reach $\varepsilon_{\rm CR} \sim 3.5\%$, as in their $\beta\sim1$ counterparts \citep[][]{orusa+23,orusa+25b}.
The combination of strongly suppressed ion acceleration and steeper-than-DSA nonthermal spectra suggests that quasi-perpendicular shocks associated with radio relics should have a very small content in relativistic ions, unlikely to produce an observable hadronic $\gamma$-ray signal.

We provide an empirical prescription for the acceleration efficiency in this regime (Equation~\ref{eq:fit}), which can be readily implemented in large-scale models of ion acceleration in ICM shocks.

We confirm that shock obliquity plays a decisive role in regulating particle injection. High-$\beta$ quasi-perpendicular shocks strongly suppress ion injection and subsequent acceleration, whereas less oblique configurations ($\vartheta_{B_n}\lesssim60^\circ$) allow more efficient suprathermal production even at relatively low Mach numbers. Moreover, such moderately oblique shocks can be effectively studied in 2D (Figure~\ref{fig:angle_eff}). The quasi-parallel counterpart of these shocks will be presented in Sharma \& Caprioli (in preparation).

Taken together, our results imply that typical radio relics, which are generally weak and quasi-perpendicular, are unlikely to contribute significantly to the cluster CR proton population. Any nonthermal proton component in galaxy clusters is therefore more likely to originate from stronger and/or quasi-parallel shocks, turbulent reacceleration, or additional physical processes not included in our laminar shock setup, such as pre-existing seed particles and upstream turbulence driven by high-$\beta$ plasma micro-instabilities \citep[e.g.,][]{schekochihin+08, reichherzer+25, diesing+25b}.

\begin{acknowledgments}
    D.C. acknowledges stimulating discussions about shocks in clusters with T.~W.~Jones, H.~Kang, D.~Ryu, and L.~Böss.
    The Authors also thank an anonymous referee for their suggestions, which led to an improved version of this manuscript.
    We thank the University of Chicago Research Computing Center for providing computational resources, as well as allocation TG AST180008 at ACCESS. This work was supported in part by NASA grant 80NSSC18K1726, NSF grants AST-2510951 and AST-2308021 to D.C. L.O. acknowledges the support of the Multimessenger Plasma Physics Center (MPPC), NSF grants PHY2206607 and PHY2206609.
\end{acknowledgments}

\clearpage

\vspace{5mm}

\appendix
\section{Simulation resolution}
\label{appendix:resolution}

Throughout this work, we adopt a numerical resolution of 8 macroparticles per cell and 8 cells per ion inertial length, $d_i$. Insufficient spatial resolution can affect the simulated shock physics, for example by producing artificial numerical cooling or by failing to resolve ion-scale instabilities such as the ion Weibel instability \citep{orusa+26}.

For the ion Weibel instability, \citet{orusa+26} recommend a spatial resolution of approximately
\begin{equation}
    N_{d_i} \simeq 9\sqrt{\frac{M_A}{30}}
\end{equation}
cells per $d_i$. For most of our simulations, the adopted resolution is close to this estimate. However, for the run with the largest Alfvénic Mach number, the recommended value is approximately twice our adopted resolution. We note that the simulations in \citet{orusa+26} were performed for low plasma beta, $\beta \sim 2$. Moreover, the ion Weibel instability is not the only process controlling particle acceleration, particularly in the high-$\beta$ shocks considered here.

To verify that our main conclusions are not sensitive to the adopted resolution, we repeated runs 5 and 15 with twice the spatial resolution, corresponding to 16 cells per $d_i$. These runs represent two opposite acceleration regimes: run 5 is a high-$\beta$, high-$M_A$ shock with a low acceleration efficiency, $\varepsilon_{\rm CR}<0.1\%$, whereas run 15 is a lower-$\beta$, high-$M_A$ shock with one of the highest measured efficiencies in our sample, $\varepsilon_{\rm CR}\sim3\%$.

\begin{figure*}[ht]
    \centering
    \includegraphics[width=\textwidth]{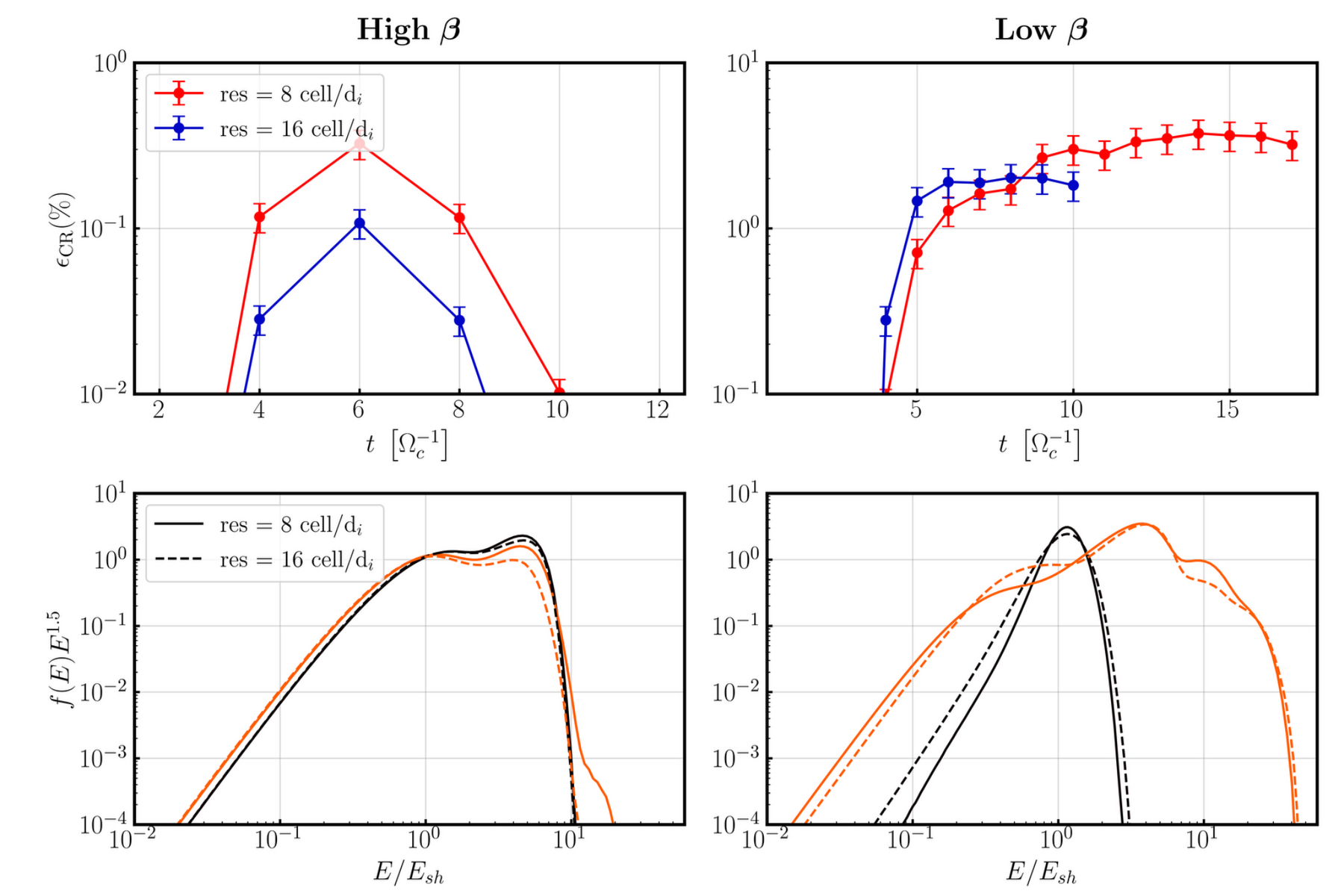}
    \caption{
    Comparison of the standard- and high-resolution simulations for runs 5
    (high $\beta$) and 15 (low $\beta$); see Table~\ref{tab:params}.
    The top row shows the time evolution of the ion acceleration efficiency,
    $\varepsilon_{\rm CR}$. The bottom row compares the ion energy spectra at
    the first analyzed timestep (black curves) and at the latest timestep
    available for the high-resolution simulations (orange curves).
    }
    \label{fig:resolution_comparison}
\end{figure*}

The results are shown in Figure~\ref{fig:resolution_comparison}. In the inefficient run, increasing the resolution causes $\varepsilon_{\rm CR}$ to decrease more rapidly with time. Nevertheless, both resolutions exhibit a trend toward vanishing acceleration efficiency at late times, further supporting the conclusion that this shock does not efficiently accelerate ions.
In the efficient run, both resolutions produce a clear population of accelerated ions and place the shock in the high-efficiency regime. Averaged over the first six analyzed timesteps for which $\varepsilon_{\rm CR}>1\%$, the standard- and high-resolution simulations yield
\begin{equation}
    \left\langle\varepsilon_{\rm CR}\right\rangle_{\rm standard}
    = (2.4\pm0.7)\%,
    \qquad
    \left\langle\varepsilon_{\rm CR}\right\rangle_{\rm high}
    = (1.9\pm0.2)\%,
\end{equation}
respectively, where the quoted uncertainties represent the temporal standard deviation. The two efficiencies are consistent with each other within the error bands.

At the latest common timestep, the difference between the two resolutions is larger, and the standard-resolution simulation subsequently approaches a higher apparent plateau. However, because the high-resolution simulation terminates at $t=10\,\Omega_c^{-1}$, a converged comparison of the asymptotic efficiencies is not possible. The corresponding ion energy spectra, evaluated at the latest available high-resolution timestep, show a somewhat less prominent suprathermal bump at higher resolution, while the high-energy tails converge to the same values. Nevertheless, both resolutions produce a clear nonthermal ion population.

These tests demonstrate that increasing the spatial resolution does not alter the qualitative conclusions of this work: the inefficient shock remains inefficient, while the shock exhibiting strong ion acceleration retains a clear nonthermal population and remains in the high-efficiency regime. Quantitatively, the resolution dependence is at the level of approximately $20\%$ for the time-averaged efficiencies.
This resolution-dependent systematic should be distinguished from the $10{-}20\%$ measurement uncertainty associated with the choice of the downstream integration region.

Because these simulations are three-dimensional, doubling the spatial resolution in all three directions increases the number of grid cells by a factor of eight. The correspondingly smaller timestep introduces an additional factor of two, resulting in an overall computational cost approximately 16 times higher. We therefore conclude that 8 cells per $d_i$ are sufficient to establish the acceleration regimes and the principal trends explored in this parameter study, while the absolute efficiencies may carry a larger resolution-dependent systematic uncertainty.

\clearpage

\bibliography{Total}
\bibliographystyle{aasjournal}
\end{document}